\documentclass{xeus}
\usepackage{graphics}
\usepackage{psfig}

\begin{document}
\onecolumn
\title{Beyond the X--ray background with {\tt XEUS}} 
\author{A. Comastri\inst{1}, P. Ranalli\inst{2,1} \and M. Brusa\inst{2,1}}
\institute{INAF -- Osservatorio Astronomico di Bologna 
 via Ranzani 1, 40127 Bologna, Italy
\and  Dipartimento di Astronomia, Universit\`a di Bologna
 via Ranzani 1, 40127 Bologna, Italy }
\maketitle

\begin{abstract}

We briefly discuss the perspectives of sensitive hard X--ray observations 
with the large collecting area X--ray telescopes of {\tt XEUS}.

\end{abstract}

\section{Introduction}

Thanks to the excellent performances of the new generation of 
X--ray satellites both in terms of spatial resolution ({\it Chandra}) and 
high energy throughput (XMM--{\it Newton}) it has been possible 
to perform the deepest X--ray observations and almost
completely resolve the 0.5--8 keV X--ray background (XRB) into single
sources (Brandt et al. 2001; Giacconi et al. 2002). 
Although at the flux limits reached by deep {\it Chandra} 
surveys a large variety of sources has been detected, 
Active Galactic Nuclei (AGN) provide the  most 
important contribution to the overall energy budget of the hard XRB.
The increasing observational evidence of a large population of obscured
sources support the prediction of AGN synthesis models for the
X--ray background (Setti \& Woltjer 1989; Comastri et al. 1995; 
Gilli et al. 2001).
At this stage it seems obvious to conclude that
there would be no other deep X--ray surveys 
if the origin of the XRB is considered a closed issue.

In the following we try to outline a few topics
which would greatly benefit from further 
observations with a large collecting area X--ray 
facility such as {\tt XEUS}. 

\section{The content of hard X--ray surveys} 

The overall status of optical follow--up observations of X--ray sources 
is usually summarized by plots of the R--band magnitude versus the 
2--10 keV X--ray flux (Figure~1).  The range of X--ray to optical flux ratios 
of spectroscopically identified AGN 
is marked with the shaded area within --1 $<$ log$(f_X/f_{opt}) <$ 1 
(Maccacaro et al. 1988). 
This trend holds for a large number 
of sources over a broad range of fluxes.
Attention should be paid to those sources which deviate from 
log$(f_X/f_{opt}) = 0 \pm 1$. 
The objects characterized by high values of 
$f_X/f_{opt}$ are optically faint, sometimes below the 
limits of deep optical images.
The spectroscopic identification of these objects 
is already challenging the capabilities of 8--10 m ground 
based telescopes. Although obscured accretion seems to provide 
the most likely explanation for their X--ray to optical 
flux ratios distribution (Comastri, Brusa \& Mignoli 2002) 
the observed properties are also consistent with those expected 
for high redshift (possibly even $z > 6$) quasars and 
cluster of galaxies at $z > 1$ . \\
A not negligible number of X--ray sources
have optical counterparts which are brighter than expected for AGN 
and lie below log$(f_X/f_{opt}) = -1$. 
The identification breakdown suggests that these objects are fairly
normal galaxies and the X--ray emission is powered by processes 
associated with star formation, even though obscured accretion 
might be responsible of the peculiar broad band properties 
of a few of them (Comastri et al. 2002a).

\begin{figure*}
\centerline{\psfig{figure=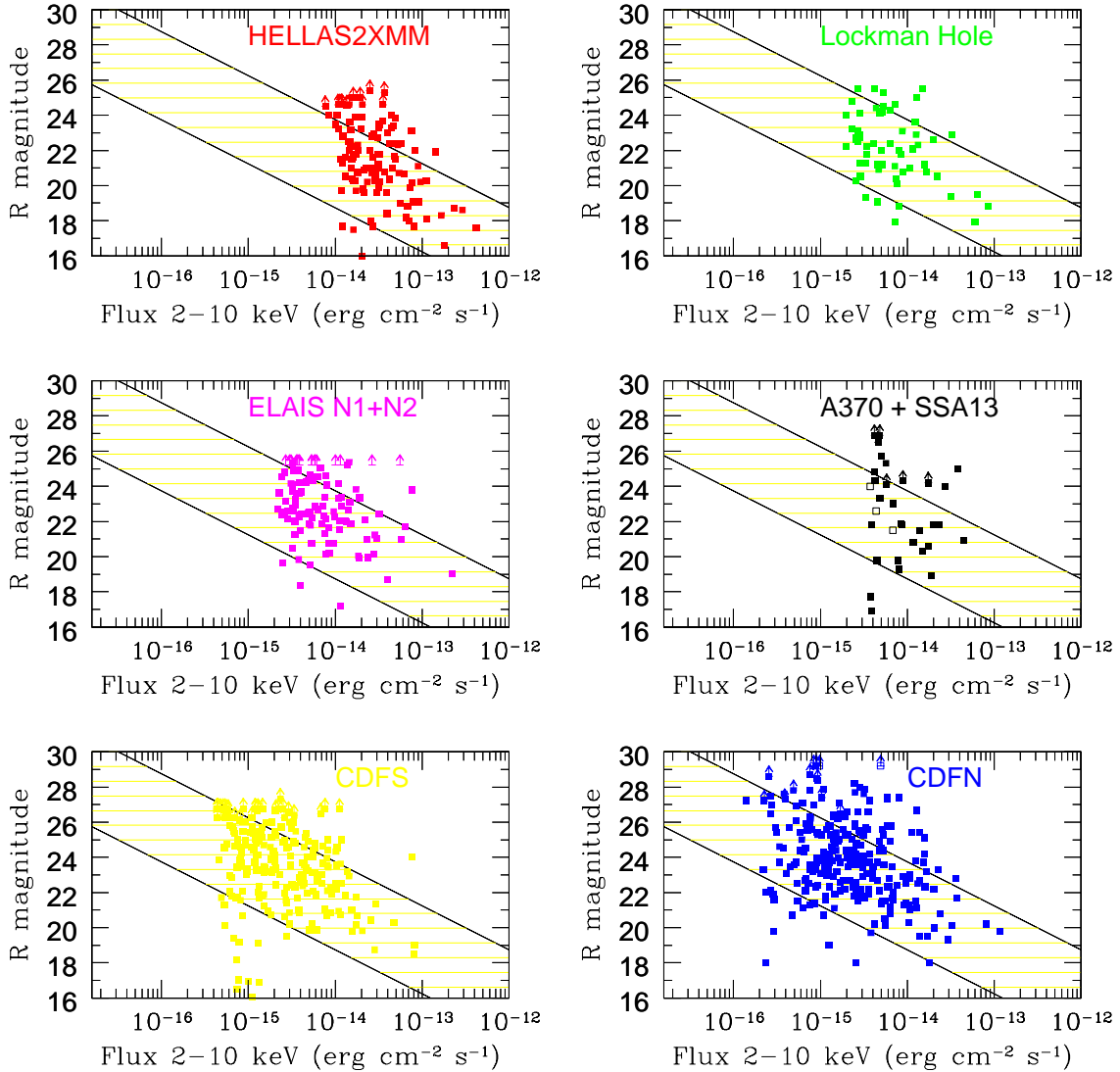,width=16cm}}
\caption[]{The 2--10 keV flux versus the R band magnitude for six
different surveys as labeled. From top left: 
the {\rm HELLAS2XMM} survey (Baldi et al. 2002), the XMM survey in the 
Lockman Hole (Mainieri et al. 2002); the {\rm ELAIS} deep {\it Chandra} 
survey (Manners et al. 2002), the combination of a few medium deep 
{\it Chandra} surveys (Barger et al. 2001), The {\it Chandra} Deep Field 
South (Giacconi et al. 2002) and the {\it Chandra} Hubble Deep Field 
North (Brandt et al. 2001). The upper (lower) solid line corresponds to 
log$f_X/f_{opt}$=1 (--1). The shaded area represents the region occupied 
by {\tt conventional} AGN (e.g. quasars, Seyferts, emission line galaxies).
}
\end{figure*}

\section{Selected outliers in the $f_X/f_{opt}$ plane}

For the purposes of the present discussion 
we have selected three different classes of hard X--ray selected sources 
which, in our opinion, deserve further investigation and would 
greatly benefit from deeper X--ray observations.
An expanded view of their X--ray and optical fluxes 
is reported in Figure 2. The red asterisks in the upper part 
of the diagram correspond to extremely red objects (EROs),
the cyan squares around log$f_X/f_{opt} \simeq -1$ 
are X--ray Bright Optically Normal 
Galaxies (aka XBONG; Comastri et al. 2002b), 
while the black circles in the lower left 
part are spectroscopically confirmed starburst galaxies.

\subsection{Extremely Red Objects} 

The bulk of the overall energy output of 
Extremely Red Objects (R--K $>$ 5 or I--K $>$ 4) 
is, almost by definition, in the near--infrared band.
The red colors are consistent with 
those expected for passively evolving galaxies, dust--enshrouded 
starburst galaxies or reddened AGN at $z \simeq $ 1. 
Recent {\it Chandra} and XMM--{\it Newton} hard X--ray observations
(Alexander et al. 2002a, Brusa et al. 2002a, Vignali et al. 2002) 
have proven to be very powerful in disentangling the various possibilities.
The properties of X--ray detected EROs are consistent with those 
of high redshift, luminous, obscured AGN. Therefore, hard X--ray selected 
EROs have the properties of the so far elusive population 
of type 2 quasars predicted by the synthesis models and responsible 
for the energetically dominant component of the XRB.
Stacking analysis of those objects which are 
not individually detected suggests that, on average, star--forming 
systems are more luminous X--ray sources than passively evolving 
galaxies (Brusa et al. 2002b).

\subsection{X-ray bright optically normal galaxies}

The excellent spatial resolution of the {\it Chandra} 
detectors allowed to unambiguosly identify an 
intriguing new class of X--ray sources 
(Hornschemeier et al. 2001, Barger et al. 2002).
They are found at moderately low redshift ($z<$ 1) and
are characterized by 
an absorption dominated optical spectrum and AGN--like hard 
X--ray luminosities ($L_{2-10} \simeq 10^{42-43}$ erg s$^{-1}$). 
The average value of their log$(f_X/f_{opt})$ distribution 
is around $-1$ with a large dispersion (Fig.~2).
An attempt to investigate their nature through a multiwavelength 
approach suggests that the putative AGN responsible
for the hard X--ray emission is completely hidden at longer wavelengths 
(Comastri et al. 2002b). It has been also suggested that the observed 
broad band properties can be explained by the presence of
an heavily obscured ($N_H > 10^{25}$ cm$^{-2}$) Compton thick 
AGN (Comastri, Brusa and Mignoli 2002).

\begin{figure*}
\centerline{\psfig{figure=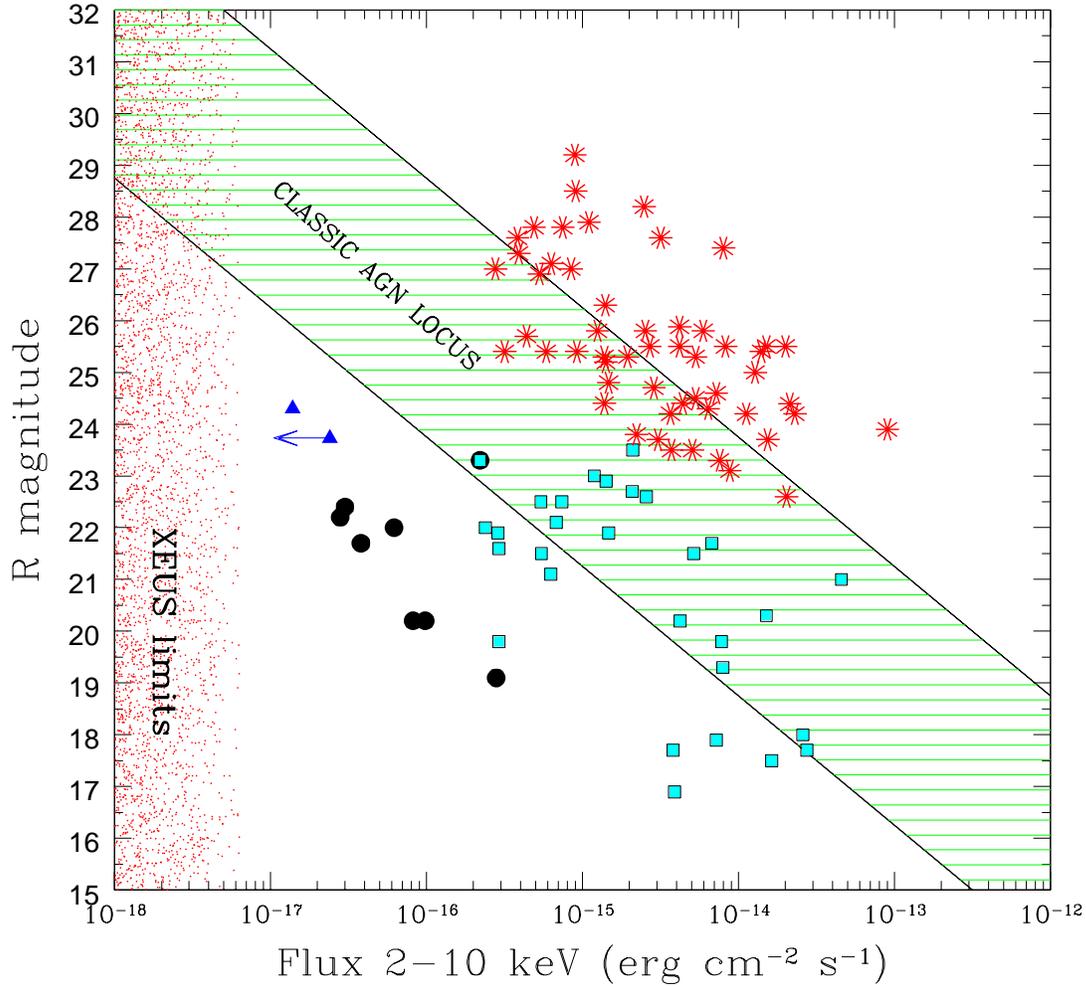,width=16cm}}
\caption[]{The same plot of Fig.~1 for a few selected examples
of recently discovered X--ray sources: EROs (red asterisks), 
{\tt XBONG} (cyan squares), and individually detected 
star--forming galaxies (black circles) in the 2 Ms {\it Chandra}
Deep Field North (Ranalli 2002). The blue triangles represent
the results of stacking analysis of those EROs associated with 
dusty star--forming system (the detection) and passively evolving 
old galaxies (the upper limit).}
\end{figure*}

\begin{figure*}
\centerline{\psfig{figure=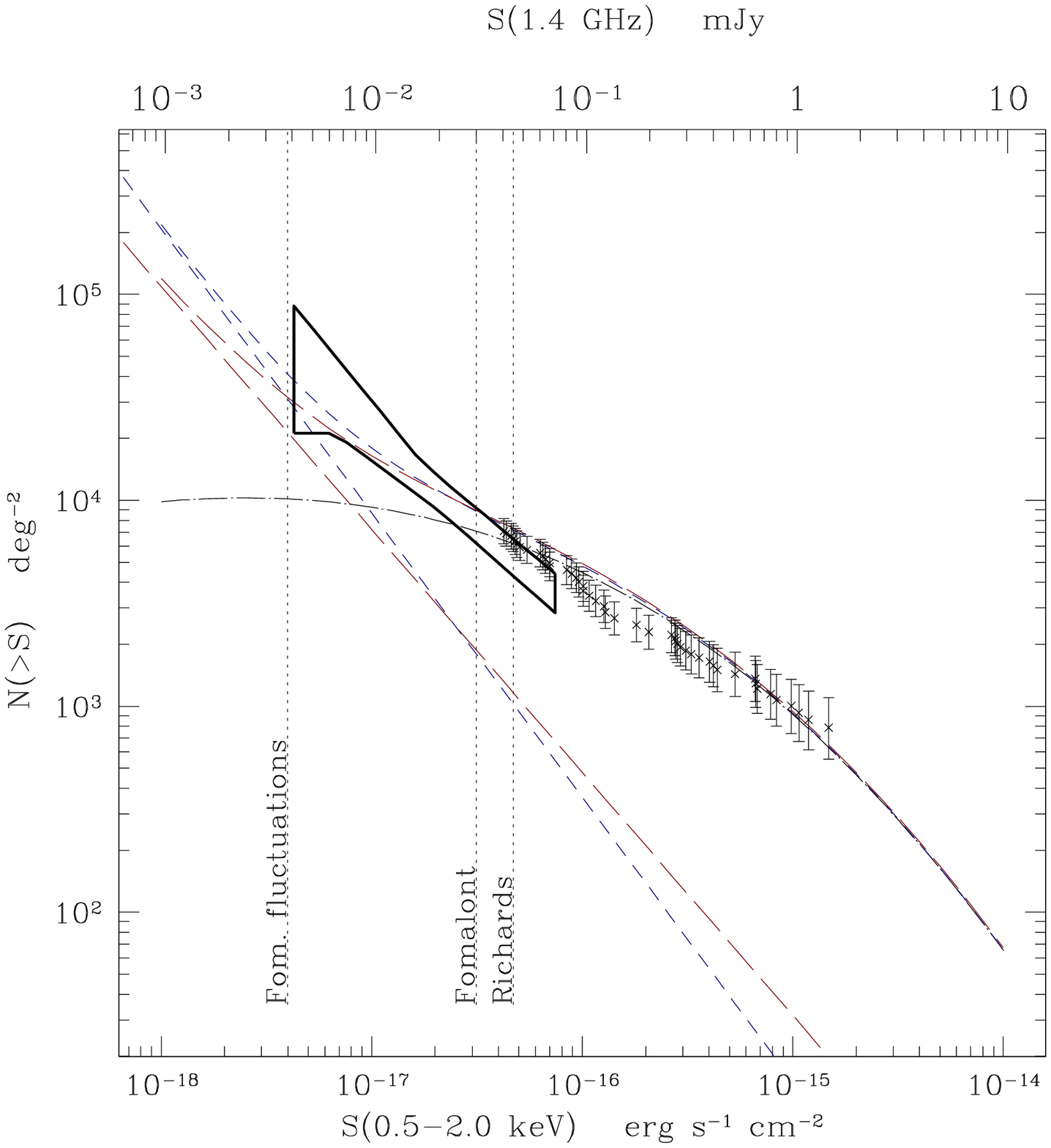,width=16cm}}
\caption[]{X-ray counts derived from deep radio Log~$N$--Log$S$.  The
blue short--dashed and red long--dashed straight lines represent X-ray
counts derived from the 1.4 GHz (Richards 2000) and 5 GHz (Fomalont et
al.~1991) Log~$N$--Log$S$ respectively.  Dots: observed X-ray number
counts in the 1 Ms {\em Chandra} HDFN survey (Brandt et al.~2001)
Horn--shaped symbols: results from X--ray fluctuation analysis
(Miyaji \& Griffiths 2002). Dot--dashed line:  number counts from AGN
synthesis models (Comastri et al.~1995).  Vertical dotted
lines: limiting sensitivities for the radio surveys. The sum of galaxies
and AGN counts is also shown.}
\end{figure*}

\subsection{The hard X--ray luminosity and the star formation rate}

It is well known that the radio and far--infrared luminosities of 
star--forming galaxies follow a tight linear relation.
Making use of a sample of nearby star--forming galaxies 
observed by ASCA and BeppoSAX it has been shown that 
a linear relation holds also between the X--ray and both 
the radio and far--infrared bands (Ranalli, Comastri \& Setti 2002).
Such a relation has been extended to $z \simeq$1  
combining the 1 Ms {\it Chandra} Deep Field North exposures 
with deep 15$\mu$m {\tt ISOCAM} (Alexander et al. 2002b)
and 1.4 GHz VLA observations (Bauer et al. 2002) 
and up to $z\simeq$ 3 via stacking analysis of 
Lyman break galaxies (Nandra et al. 2002).
These results imply that the hard X--ray emission can be used as 
an absorption independent indicator of the star formation rate.\\
Moreover it is possible to compute the expected X--ray counts
of star--forming galaxies from the Log$N$--Log$S$ measured in 
deep radio surveys (Fig.~3). The number  counts of 
star--forming galaxies begin to overcome AGN counts at fluxes of the 
order of 10$^{-17}$ erg s$^{-1}$ cm$^{-2}$. Interesting enough 
the predictions are fully consistent with the constraints 
from fluctuation analysis in deep {\it Chandra} fields
(Miyaji \& Griffiths 2002) and with the number counts expected 
by the X--ray binaries resulting from the peak in the cosmic star 
formation rate at $z >$1 (Ptak et al. 2001).

\section{Perspectives for {\tt XEUS} deep fields}
The foreseen {\tt XEUS} capabilities in terms of collecting area 
will allow to obtain 
(with exposure times ranging from 100 ks to 1 Ms)
good quality X--ray spectra 
for almost all the X--ray sources discovered in the deepest 
{\it Chandra} and XMM--{\it Newton} surveys.
Given that most of these sources are likely to be the obscured 
AGN responsible for the bulk of the 
hard X--ray background, it will be possible to completely 
characterize the cosmic history of accretion powered sources.
It is also worth noticing that the detection of 
iron K$\alpha$ features could provide the only 
method (besides spectroscopy with  30--100 m  optical telescopes
and the photometric technique)  
to obtain a redshift estimate for those sources 
with extremely faint optical counterparts.
The detection of X--ray sources will be pushed down to limiting 
fluxes of the order of a few $10^{-18}$ erg s$^{-1}$ cm$^{-2}$.
At this level starburst galaxies will be detected in the X--ray band
to at least $z\simeq$3, allowing to obtain an independent 
constraint on the cosmic star formation history to be compared 
with those obtained at longer wavelengths, and to investigate 
the connection with the onset and fueling of massive black holes.

\begin{acknowledgements}
 This research has been partially supported by ASI contracts
I/R/113/01; I/R/073/01 and by the MIUR grant Cofin--00--02--36. We thank
G.G.C. Palumbo for useful comments.
\end{acknowledgements}

\end{document}